# A Method for Determining the Transient Process Duration in Dynamic Systems in the Regime of Chaotic Oscillations

A. A. Koronovskii[a,*], A. V. Starodubov[a], and A. E. Hramov[b,**]

[a] *Saratov State University, Saratov, Russia*
\* *e-mail: alkor@cas.ssu.runnet.ru*
[b] *State Scientific Center "College", Saratov, Russia*
\*\* *e-mail: aeh@cas.ssu.runnet.ru*

**Abstract**—We describe a method for determining the transient process duration in a standard two-dimensional dynamic system with discrete time (Eno map), occurring in the regime of chaotic oscillations.

In most investigations devoted to nonlinear dynamic systems exhibiting complicated behavior, the attention of researchers is concentrated on the established regimes (that can be both periodic and chaotic). The initial (transient) stage of the process studied is frequently considered as insignificant and rejected, the analysis being mostly devoted to determining the characteristics (attractor dimensions, Lyapunov exponents, etc.) of established oscillation regimes, bifurcations arising upon variation of the control parameters, scenarios of the transitions to chaos, and so on. However, the transient processes also obey certain laws [1] and sometimes can provide important information about the whole system and its dynamics [2].

In studying a transient process, an important point is to determine (to within a preset accuracy $\varepsilon$) the moment when this process comes to an end and the system attains (with the same precision $\varepsilon$) an asymptotic state. In cases when the asymptotic state of a dynamic system of finite dimension is periodic (immobile stable point, periodic cycle), there is a quite simple and sufficiently effective algorithm of determining the time instant corresponding to the system reaching this state [1, 3]. The essence of this method can be formulated as follows: for a dynamic system with discrete time, (i) each element of a periodic attractor composed of a finite small number of points is surrounded by a circle of radius $\varepsilon$ and (ii) the time when the imaging point falls within one of these $\varepsilon$-neighborhoods is monitored. (Note that any dynamic flow system of finite dimension can be reduced to a map by using the Poincaré section procedure [4, 5]).

However, the above approach is inapplicable to dynamic systems occurring in a chaotic state. The main difficulty is related to the fact that the number of points belonging to the chaotic attractor is infinite. This leads to difficulties in determining the necessary number of attractor points by which the transient process duration will be evaluated: this number proves to be very large and strongly depends on the accuracy $\varepsilon$ with which the process duration has to be determined [3]. Accordingly, the time required for trials involving the attractor elements substantially increases.

This study aimed at the creating an effective method for determining the transient process duration in a standard two- dimensional dynamic system with discrete time,

$$\mathbf{x}_{n+1} = \mathbf{F}(\mathbf{x}_n), \qquad (1)$$

occurring in a chaotic regime. We have selected a standard two-dimensional dynamic system with discrete map of the type known as the Eno map [6, 7],

$$x_{n+1} = \lambda x_n(1 - x_n) + b y_n, \quad y_{n+1} = x_n, \qquad (2)$$

where $\lambda$ and $b$ are the control parameters determining the oscillation mode. Below, we will consider the case of $\lambda = 2.5453$ and $b = 0.5188$, which corresponds to a chaotic oscillation regime in system (2).

By the transient process duration in a system with discrete time is implied an interval of $K$ discrete time units, after elapse of which the imaging point in the phase space attains an attractor to within a preset accuracy $\varepsilon$. In other words, for all $n > K$

$$\left\| \mathbf{x}_n - \mathbf{x}_j^0 \right\| < \varepsilon, \qquad (3)$$

where $\mathbf{x}_n$ are the points of time series of the given dynamic system, $\mathbf{x}_j^0$ ($j = \overline{1, \ldots, M}$) are the elements of a chaotic attractor, and $\varepsilon$ is the preset accuracy with which the transient process duration has to be determined. Formally speaking, a transient process is infinitely long: the imaging point asymptotically approaches the attractor as $t \longrightarrow \infty$, never reaching it for a finite time. On the other hand, regions corresponding to an established regime and the transient process can usually be separated in the time series obtained by the experiments under natural conditions and by numerical modeling of the behavior of various systems [8, 9]. The transient process duration depends on the control parameters, initial conditions, and the preset uncertainty $\varepsilon$ within which the imaging point reaches the attractor.

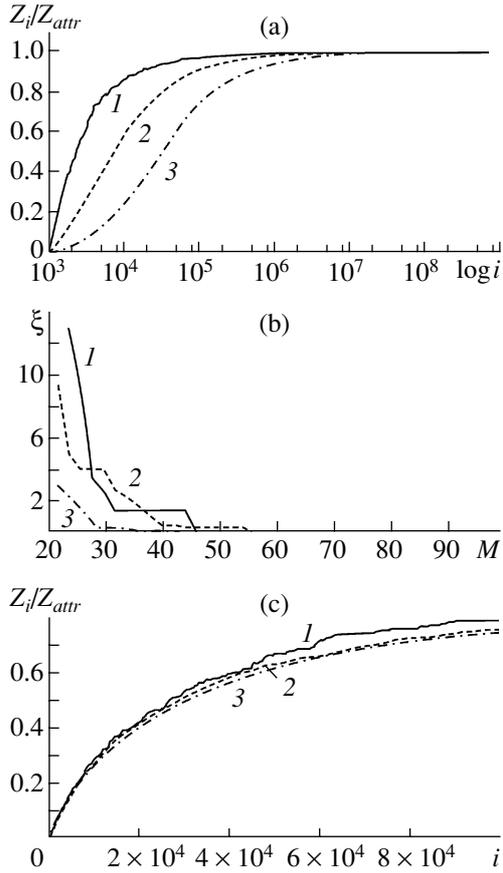

**Fig. 1.** Plots of (a) the number of attractor cells $Z_i$ (normalized to the total number of cells $Z_{attr}$ found for a preset accuracy $\varepsilon$) versus the number of iterations $i$, (b) percentage of errors $\xi$ in determining the transient process duration versus the percentage $M$ of attractor cells used for this determination, and (c) the normalized number of attractor cells $Z_i$ versus the number of iterations (normalized to $\mu$) for various values of the accuracy: $\varepsilon = 10^{-2}$ (*1*), $10^{-5/2}$ (*2*), and $10^{-3}$ (*3*). Control parameters: $\lambda = 2.5453$ and $b = 0.5188$. Initial conditions: $x_0 = 0.1$ and $y_0 = 0.1$.

The proposed method is essentially as follows: the region of initial conditions containing the attractor is covered with a lattice of period (step) $\varepsilon$. Then an initial condition $\mathbf{x}_0$ (i.e., an initial point in the attraction basin) is selected and the system (1) is iterated a sufficiently large number of times $N$ (definitely greater than the maximum transient time duration $T_{max}$). In the course of this iteration process, the imaging point travels over the lattice following the attractor and passing through various lattice cells. Beginning with the discrete time instant $i = T_{max} + 1$, we count the number of cells visited by the imaging point during the iterative procedure. These cells, covering the attractor points $\mathbf{x}_j^0$ will be referred to as the attractor cells. After termination of the iterative procedure, we obtain a basis array of such attractor cells. Once the cells belonging to the attractor are known, the transient process duration can be determined as follows: if the initial point $\mathbf{x}_0$ falls within an attractor cell upon iteration for $K$ discrete time units, the transient process duration determined for the given initial condition to within a preset accuracy $\varepsilon$ is $K$. Obviously, correct determination of the transient process duration requires finding possibly large number of the attractor cells.

Since it is considered that, past the transient process of $K$ discrete time units, the imaging point reaches a chaotic attractor, the condition (3) must be satisfied for all $i > K$. Theoretically, a situation is possible whereby the imaging point $\mathbf{x}_i$, approaching one of the chaotic attractor points $\mathbf{x}_{j\,0}^0$ to a distance below $\varepsilon$ upon the $i$th iteration, does not in fact reach the attractor: condition (3) fails to be valid upon the $(i+1)$th iteration and no one point of the chaotic attractor falls within the $\varepsilon$-neighborhood of point $\mathbf{x}_{i+1}$. In this case, completion of the transient process is verified by checking the condition (3) for $m$ sequential control points $\mathbf{x}_{i+1}, \ldots, \mathbf{x}_{i+m}$. However, this verification procedure was shown [3] to be unnecessary, only significantly increasing the time required for correctly determining the transient process duration by trying a large number of points of the chaotic attractor.

In our study, the iterative procedure included $N = 5 \times 10^8$ steps and the maximum iterative process duration was $T_{max} = 10^3$. The so large number of iterations was necessary to provide for a more complete basis array, offering a good approximation to the true chaotic attractor and possessing all its essential properties. A numerical experiment showed that, for $\varepsilon = 10^{-2}$ and the corresponding number of iterations, the number of attractor cells is $Z_{attr} = 1228$; for $\varepsilon = 10^{-5/2}$, $Z_{attr} = 5622$; and for $\varepsilon = 10^{-3}$, $Z_{attr} = 25199$.

Now let us consider the dependence of the number of attractor cells $Z_i$ on the number of accomplished iteration steps $i$ for which the attractor cells were found at a given $\varepsilon$ (Fig. 1a). As can be seen from the character of these curves, the rate of increase in the number of attractor cells is sufficiently large in the initial stage of the time series. As the number of iterations increases, this rate drops and eventually the curves exhibit saturation. This behavior indicates that further iteration process is unexpedient: new cells are very rarely (or not at all) added to the attractor. Obviously, the higher the accuracy (the lower the error) of determination of the transient process duration, the greater number of iterations $i$ has to be performed in order to provide that the $Z_i$ curve reaches the level of saturation for $Z_{attr}$.

Once it is known which cells on the $(x, y)$ plane belong to the chaotic attractor, we can study the depen-dence of the transient process duration for the Eno map with various initial conditions $(x_0, y_0)$ on the number of attractor cells $Z_{attr}$ found. Figure 1b shows the plots of the percentage of errors in determining the transient process duration versus percentage $M$ of the basis array cells used for this determination. Here, by errors are implied the situations when the transient process duration determined for the initial condition $(x_0, y_0)$ using 100% of the basis array cells differs from that determined using only $M$ [%] of such cells. The initial conditions were represented by an array of $10^3$ random points $(x, y)$. Obviously, the smaller the number of the attractor cells used for determining the transient process duration, the greater the percentage of errors. As can be seen from Fig. 1b, there is a certain value of $M$ < 100%, above which the number of errors in determining the transient process duration is minimum.

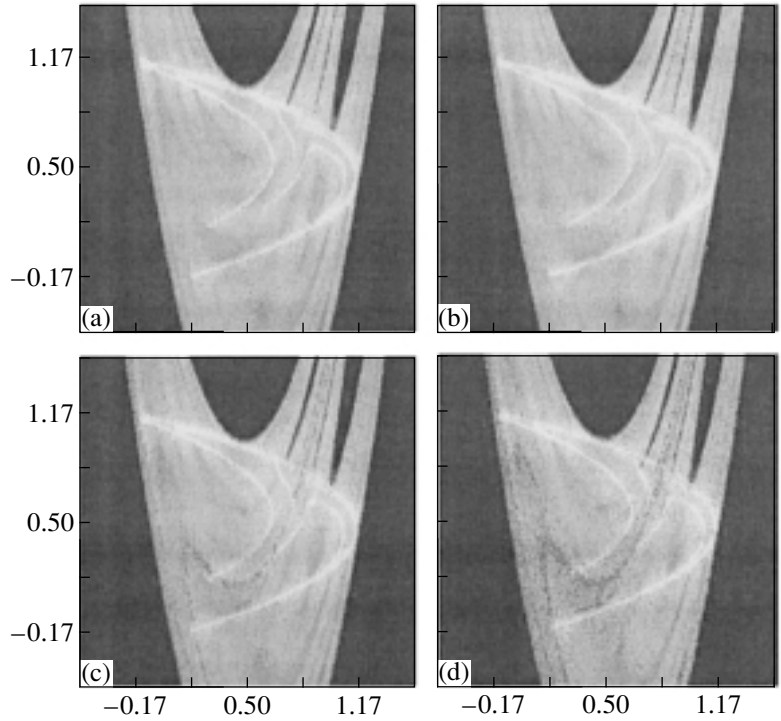

**Fig. 2.** Gray gradation maps showing the transient process duration as a function of the initial conditions for the Eno map (2), calculated using various numbers of the attractor cells $M$ = 100% (a), 90% (b), 80% (c), and 60% (d) for the control parameters: $\lambda$ = 2.5453 and $b$ = 0.5188 and the accuracy $\varepsilon = 10^{-3}$. The maximum transient process duration is $T$ = 20. Black areas correspond to the initial conditions from which the imaging point goes to infinity.

Figure 2 presents the maps of the transient process duration as a function of the initial conditions $(x_0, y_0)$ plotted using the proposed method for various numbers of attractor cells. These maps were constructed as follows: each point of the $(x, y)$ plane was considered as the initial condition, the system with this initial condition was characterized by the transient process duration, and the given point was painted with the corresponding color gradation. The greater the transient process duration, the darker the gray color gradation for this point. In this way, the chaotic attractor was painted white. As can be seen from Fig. 2, a decrease in the number of basis array cells used for determining the transient process duration leads to the appearance of darker (or even black) points. This is explained by the fact that the transient process duration is incorrectly calculated for some initial conditions.

Thus, the above data indicate that the transient process duration can be determined using a fraction of the basis array cells, rather than the whole array. With this fraction, a relatively small number of iterations $i$ can be accomplished. This number depends on the accuracy $\varepsilon$ with which the transient process duration has to be determined. In connection with this, it is of interest to study a relation between the capacitive dimension $D_0$ of a chaotic attractor [10] and the number of iterations $i$ of map (2) necessary for correctly determining the transient process duration to within a preset accuracy $\varepsilon$. In other words, the question is whether we can find, proceeding from the known attractor dimension $D_0$, a scaling factor that would allows us to determine the number of iterations necessary for correctly calculating the transient process duration with a higher accuracy (or lower $\varepsilon$). In order to elucidate the above question, we have calculated the attractor dimension and obtained $D_0 = 1.31$. For two values of the accuracy, $\varepsilon_1$ and $\varepsilon_2$ such that $\varepsilon_2 = \varepsilon_1/a$, the numbers of the attractor cells are $N_1(\varepsilon) \sim \varepsilon_1^{-D_0}$ and $N_2(\varepsilon) \sim \varepsilon_2^{-D_0}$. Thus, the scaling factor is determined as $\mu = N_2/N_1 = (\varepsilon_2/\varepsilon_1)^{-D_0} = a^{D_0}$. For $a = \sqrt{10}$, we obtain $\mu = 10^{D_0/2} \approx 4.54$. Figure 1c shows the plots of $Z_i/Z_{attr}$ for various values of $\varepsilon$, with the number of iterations normalized to the above scaling factor $\mu$. As can be seen from Fig. 1c, the curves coincide with each other to within a rather high precision. Therefore, once the necessary number of iterations $N_1$ for correctly determining the transient process duration with a preset accuracy $\varepsilon_1$ is known, it is possible to calculate the value of necessary iterations $N_2$ for any other $\varepsilon_2$.

Thus, we have proposed an effective and rapid method for determining the transient process duration in a two-dimensional dynamic system with discrete time, occurring in the regime of chaotic oscillations. This approach is also applicable to systems with continuous time, which can be reduced to maps via the Poincaré section method.

**Acknowledgments.** This study was supported by the Russian Foundation for Basic Research (project nos. 01-02-17392 and 00-15-96673) and by the Scientific-Education Center "Nonlinear Dynamics and Biophysics" at the Saratov State University (Grant REC-006 from the US Civilian Research and Development Foundation for Independent States of the Former Soviet Union).